# Mechanical aspects of Near-Infrared Imager Spectrometer and Polarimeter

Prashanth Kumar Kasarla[1], Pitamber Singh Patwal[1], Hitesh Kumar L. Adalja[1], Satya Narain Mathur[1], Deekshya Roy Sarkar[1], Alka Singh[1], Archita Rai[1,2], Prachi Vinod Prajapati[1], Sachindra Naik[1], Amish B. Shah[1], Shashikiran Ganesh[1], Kiran S. Baliyan[1]

[1]Physical Research Laboratory, Ahmedabad, India
[2]Indian Institute of Technology, Gandhinagar, India

## ABSTRACT

Near-infrared Imager Spectrometer and Polarimeter (NISP) is a camera, an intermediate resolution spectrograph and an imaging polarimeter being developed for upcoming 2.5m telescope of Physical Research Laboratory at Mount Abu, India. NISP is designed to work in the Near-IR (0.8-2.5 micron) using a H2RG detector. Collimator and camera lenses would transfer the image from the focal plane of the telescope to the detector plane. The entire optics, mechanical support structures, detector-SIDECAR assembly will be cooled to cryo-temperatures using an open cycle Liquid Nitrogen tank inside a vacuum Dewar. GFRP support structures would be used to isolate cryogenic system from the Dewar. Two layer thermal shielding would be used to reduce the radiative heat transfer. Molecular sieve (getter) would be used to enhance the vacuum level inside Dewar. Magnet-reedswitch combination are used for absolute positioning of filterwheels. Here we describe the mechanical aspects in detail.

## 1. INTRODUCTION

This document describes the mechanical design aspects of the NISP instrument. Instrument is designed to work in wavelength range (0.8-2.5 micron). All the objects at room temperature emit significant blackbody radiation in the above wavelength range. Detector, Optics and the support mechanical structure are all cooled down to cryogenic temperature to decrease the background radiation. Instrument mechanical structure is designed to enclose Optics, H2RG Detector and Cryogenic SIDECAR electronics subsystems. To achieve cryogenic temperature at the above subsystems, high vacuum conditions ($5 \times 10^{-6}$ mbar) are maintained. Optical and detector assemblies are mounted on a cold plate. A liquid nitrogen tank cools cold plate to cryogenic temperature (77K). High vacuum facilitates stable operating temperature by preventing heat transfer by conduction and convection. Thermal shields are introduced between Dewar and cold surfaces to reduce heat transfer by radiation. A scaled down version (Test Dewar) of NISP Dewar is developed to test all the subsystems.

**Design requirements:**

| | |
|---|---|
| Detector operating temperature | : 77±0.1 K |
| Max cooling rate/Warm-up rate | < 1K/min |
| Vacuum | : $5 \times 10^{-6}$ mbar |
| LN2 hold time | : 32hrs |
| Cold plate temperature | : 80K |
| Instrument mass | : 400Kg |

## 2. MECHANICAL LAYOUT

NISP optics is divided into different sub assemblies for mechanical simplicity and easy integration. All the sub-assemblies can be independently assembled and disassembled on the cold plate connected to LN2 tank as shown in Figure 1b. All the sub-assemblies are enclosed inside cold shield directly connected to cold bench to avoid any direct radiation load from the Dewar surface. Vapor cooled warm shield reduces the effective radiation load on the LN2 tank. This has direct implication on LN2 hold time.

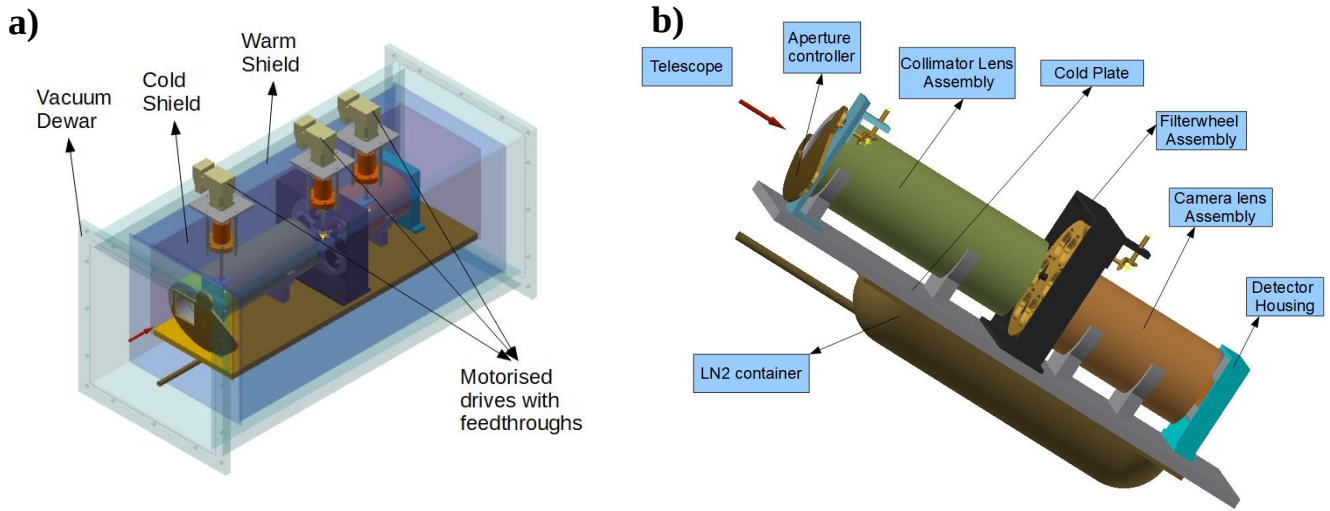

Figure 1: Instrument schematic

## 2.1 Filterwheel assembly

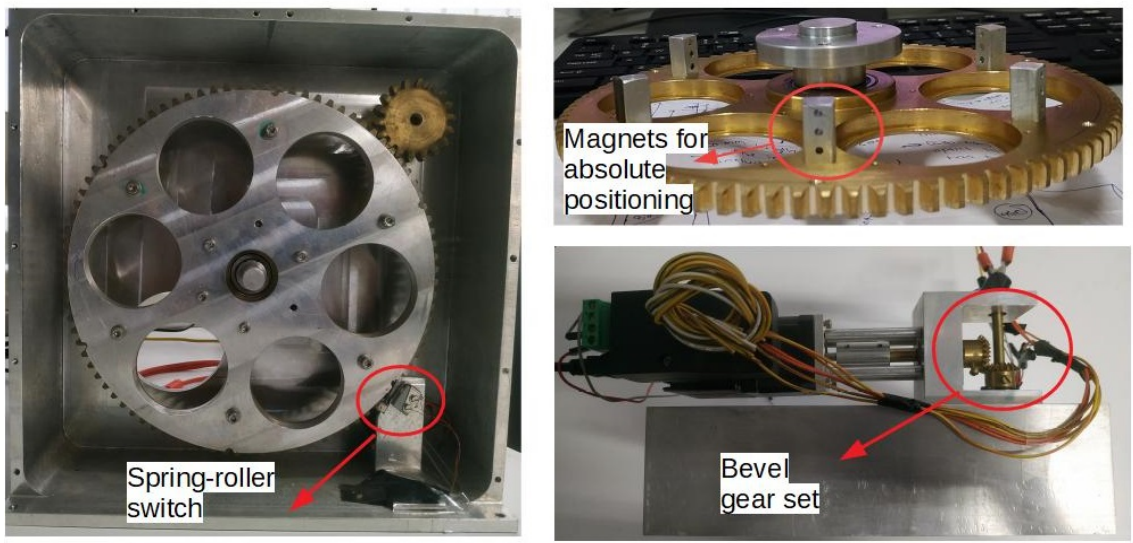

Figure 2: Filterwheel assembly

Filterwheel assembly has 3 independently driven filterwheels. Each filterwheel has 6 filter positions. Smart stepper motors are being used for motion control. Since these motors cannot be used at cryogenic temperatures and high vacuum conditions, they will be mounted outside the Dewar in ambient conditions. Rotary feedthrough would be used to transmit the motion to the filterwheels inside the Dewar. Mechanisms are equipped with magnet-reedswitch combination for absolute positioning and spring-roller switch for locking. The drive mechanism consists of bevel gear set and spur gears set. One of spur gears is the filterwheel. All the bevels, pinion gear and the filter wheel are made of brass material and SS 304 shafts are used. Significant backlash is introduced between gears to reduce the interference.

## 2.2 ROIC-Cryo SIDECAR assembly:

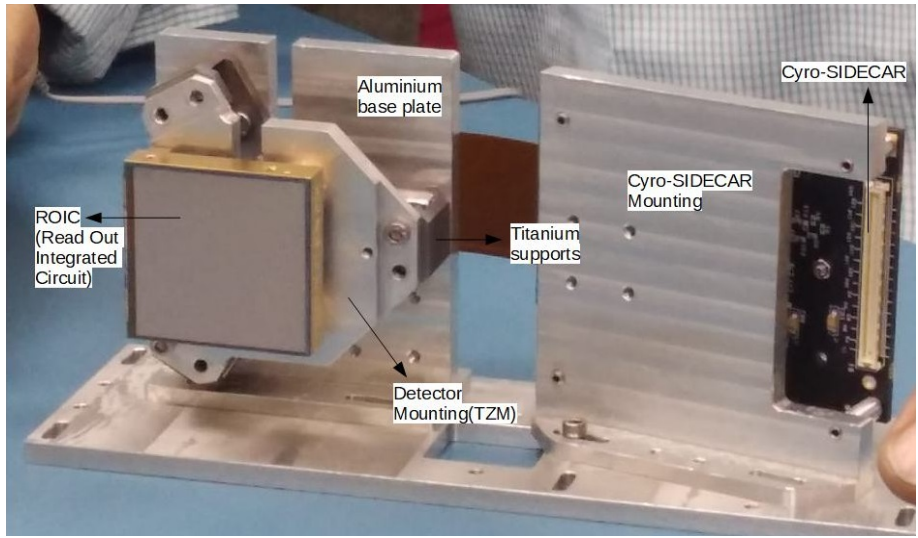

Figure 3: ROIC-Cryo SIDECAR assembled in 10K class cleanroom in PRL, Ahmedabad.

ROIC is fixed on mounting base made of TZM alloy. Having same CTE as that of detector encloser reduces the thermal stress on the detector. Titanium supports legs connects TZM base structure to aluminium base plate. Flex cable from the detector connects to the cryo-SIDECAR.

## 3. THERMAL DESIGN

Conduction and radiation thermal load on LN2 tank from different sources is listed below:
1. Conduction from G10 supports : 4W
2. Radiation from optical window : 3.5W
3. Conduction through motor drive shafts : 1.25W
4. Conduction through Wiring and Flex cable : 0.75W
Net heat transfer from all the above : 9.5W

Based on the mechanical model shown in the earlier section, the dimensions of cold surface and Dewar inner surface are fixed and radiation shields are introduced in different configurations. The radiation load in each configuration is calculated as shown below.
a) Thermally isolated radiation shield

Radiation heat load : 14.11 W
Temperature of warm-shield($T_3$) : 259K
Net heat load (H) : 14.11 + 9.5 = 23.61W
LN2 requirement for 24hrs = H x 86400/$L_{N2}$
= 10.25kg = 12.81 litres
LN2 tank size(hold time of 32 hrs) = 13.66 kg = 17 litres

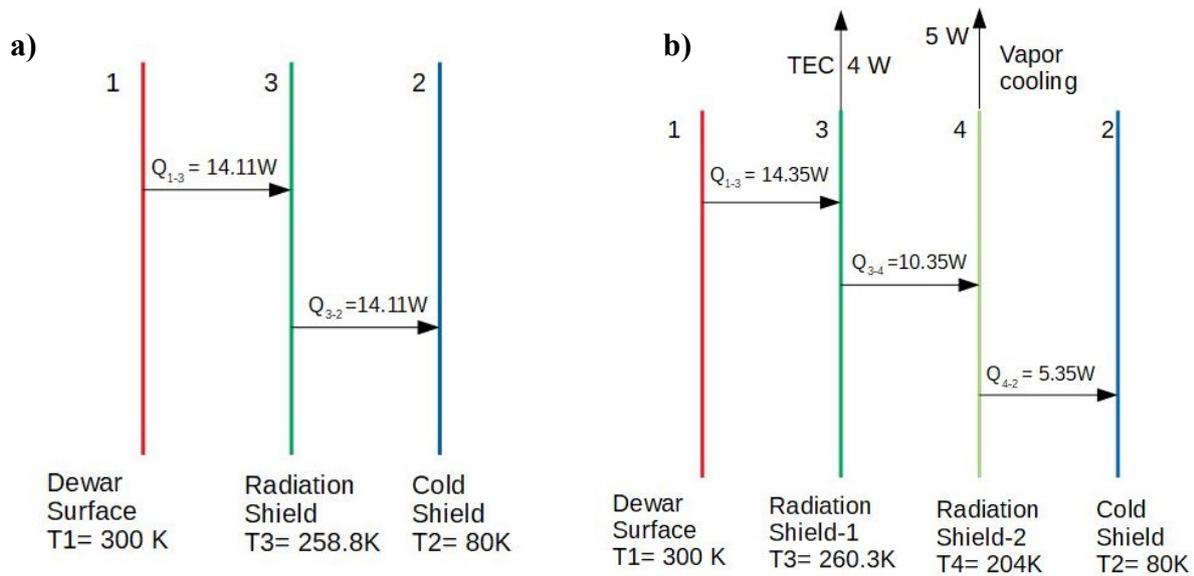

Figure 4: a) Thermally isolated radiation shield b) TEC cooled and vapor cooled radiation shield

b) TEC cooled first shield and vapor cooled second shield

Radiation heat load                          : 5.35W
Net heat load(H)                             : 5.35+ 9.5    = 14.85W
Temperature of shield-1($T_3$)               : 260.3K
Temperature of shield-2($T_4$)               : 204K
LN2 requirement for 24hrs                    = H x 86400/$L_{N2}$
                                             = 6.44kg = 8 litres
LN2 tank size(hold time of 32 hrs) =  8.59kg =10.74 litres

$L_{N2}$ – Latent heat of vaporisation of nitrogen  = 199KJ/kg

Based on above calculations, radiation heat load on the system can be reduced from 14.11W to 5.35W (64% reduction).

## 4. TEST DEWAR SETUP

Test Dewar is designed to test full size Detector-Cryo SIDECAR assembly and filterwheel assembly. Modified IR labs ND-5 Dewar and LN2 tank is used with the Aluminium Dewar (LXBXH =300mmX300mmX200mm) manufactured at PRL workshop(shown in Figure 5). Test Dewar has a cold plate that can accommodate above sub-systems. The cold plate is connected to 1.2 litre LN2 tank. Similar to full size model, it also has a cold shield, vapor cooled radiation shield. Single O-ring seals are used at all the joints on the Dewar. Viton O-rings are used for low outgassing. Standard activated charcoal is used to maintain vacuum for long term use. Dewar has 6 feedthroughs (2 electrical, 1 for vacuum gauge, 1 for pumping port, 1 for charcoal getter). Full range Active pirani/Cold cathode gauge is used to measure pressure inside the Dewar. 5 silicon diodes are used to measure temperatures at various locations.

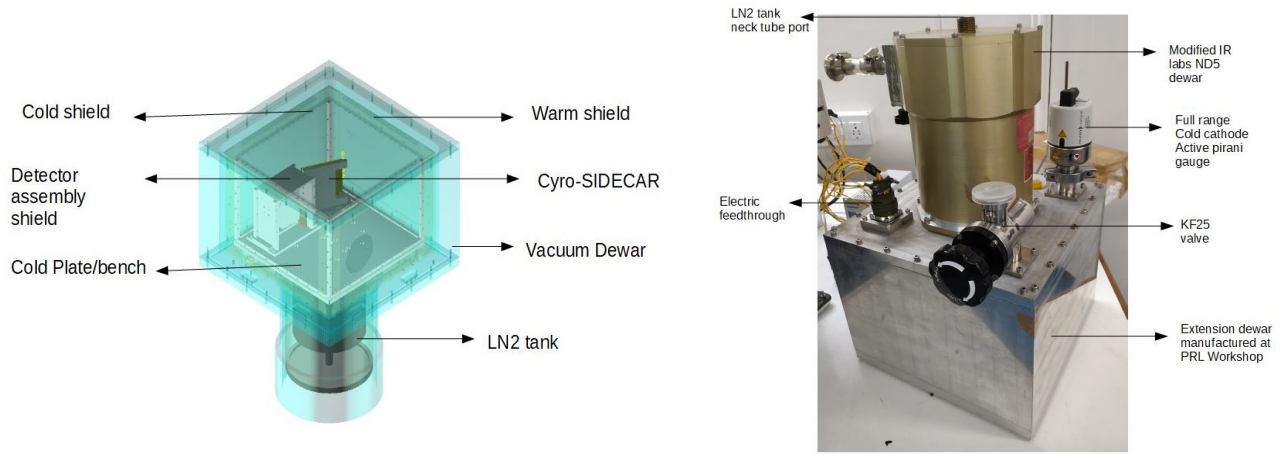

Figure 5: a) Test Dewar schematic    b) Fully assembled Test Dewar

### 4.1 Leak rate estimation

#### 4.1.1 Pressure Profile test

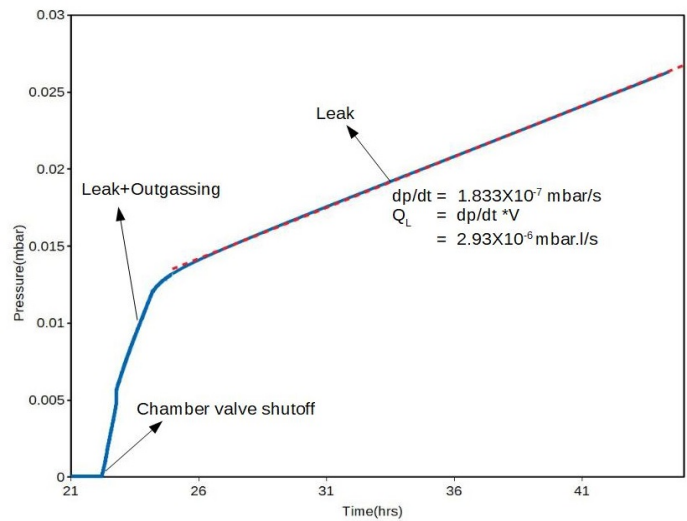

Figure 6: Test Dewar pressure profile

Vacuum Dewar was pumped using Pfieffer Eco 80 turbo pumping station. The pressure in the chamber got stabilised at $4.9 \times 10^{-5}$ mbar. Chamber valve connecting to the pump is closed. The pressure rise in the chamber in shown in the Figure 6. The starting rise in pressure is the combination of leak and outgassing. Linear steady rise in pressure is the contribution of leak.

From the plot, dp/dt is calculated to be: $1.833 \times 10^{-7}$ mbar/s.

$Q_L(air) = (dp/dt) \times V = 2.93 \times 0^{-6}$ mbar.l/s

V – Dewar Volume (excluding the space occupied by assembly parts inside dewar) = 16 litres

### 4.1.2 Mass spectrometer leak test

Leak rate of an aperture is a measure of number of gas particles going into the chamber through it. For a given aperture or point of leak.

$Q_L(gas) \propto P(Y/MW)^{0.5}$

P – partial pressure of gas
Y – ratio of specific heats
MW – molecular weight of gas

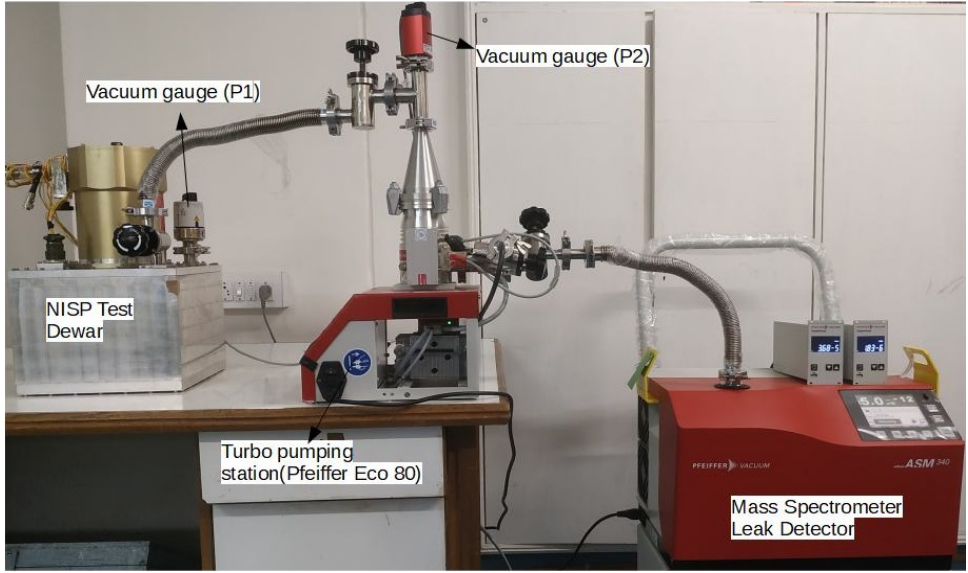

Figure 7: Vacuum Dewar leak test using Mass spectrometer leak detector

Leak detector measures helium leak rate $Q_L(helium)$.

$Q_L(air) = Q_L(helium) \times P(air)/P(helium) \times 0.34$

In air, $P(helium) = 5 ppm = 5 \times 10^{-6} P(air)$

$Q_L(air) = Q_L(helium) \times P(air)/P(helium) \times 0.34$

$Q_L(air) = 6.8 \times 10^4 \times Q_L(helium)$

As shown in the Figure 7, Leak detector is connected to the outlet of the turbo pump to increase the sensitivity of the instrument.

Helium Leak rate of the Test Dewar $Q_L(helium)$ = $2.9 \times 10^{-11}$ mbar.l/s.
Leak rate of the Test Dewar $Q_L(air)$ = $6.8 \times 10^4 \times Q_L(helium)$
= $1.97 \times 10^{-6}$ mbar.l/s

Gas permeation from O-rings seals is $2 \times 10^{-6}$ mbar.l/s. Leak rates from both methods are similar to the gas permeation from O-ring seals. From this, we can state that system is leak tight.

## 4.2 Charcoal getter testing

Two tests were conducted in the Test Dewar with and without getter to experimentally understand the working of charcoal getter. Decrease in pressure as LN2 is filled is due to the condensation of water vapor on the cold surfaces of LN2 tank and cold surfaces. Pressure in the Dewar with getter is $2 \times 10^{-6}$ mbar (less than required $5 \times 10^{-6}$ mbar). It can be observed from Figure 8, that getter is not only used to maintain vacuum for long durations, it also improves vacuum level.

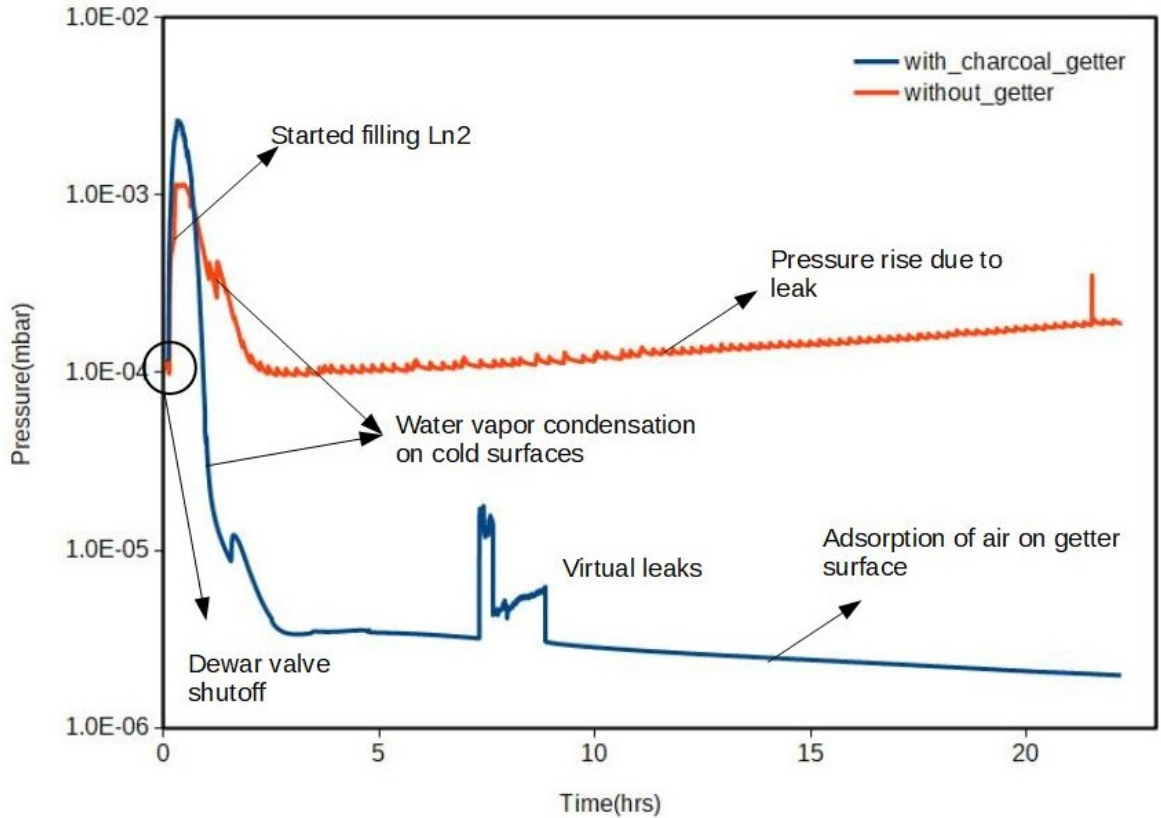

Figure 8: Pressure profiles in the chamber with and without getter

## 4.3 Thermal Analysis (Test Dewar)

### 4.3.1 Effect of vapor cooling of radiation shield.

Two experiments are carried out to find the effect of vapor cooling. In first experiment, the shield is thermally isolated and temperature measured is 259K. Radiation heat load in thermally isolated radiation shield configuration is 2.62W. In second experiment, the shield is connected to the LN2 tank neck tube. Temperatures measured at two positions on the shield (near the connection with LN2 tank tube, one on the other end) are 238K and 243K. Taking the average temperature of the shield as 241K. Radiation heat load in vapor cooled shield configuration is 1.96W. There is a 25% decrease in radiation heat load due to vapor cooling. Dry nitrogen vapors removed 1.63W of heat from thermal shield.

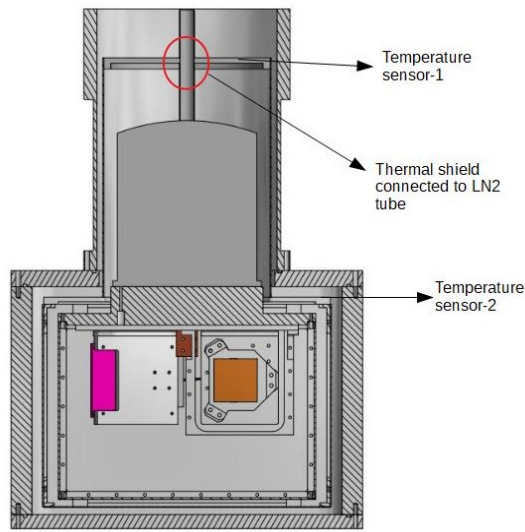

Figure 9: Sectional view of the test Dewar showing the position of temperature sensors

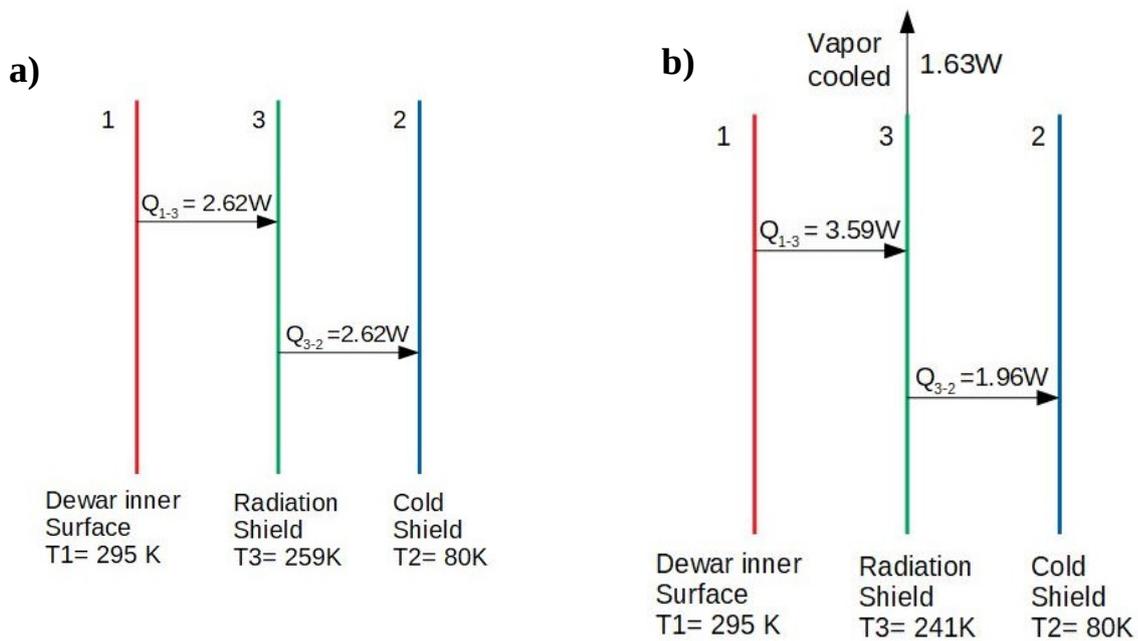

Figure 10: a) Thermal analysis of thermal isolated shield   b) Thermal analysis of vapor cooled shield

**4.3.2 Efficiency of vapor cooling system**

Experimental LN2 hold time of the vapor cooled system is 23.3hrs.  Mass evaporated per unit time is $1.14 \times 10^{-5}$ kg/s. The maximum amount of heat that can be extracted from these cool nitrogen vapors is 2.517 W.    From previous analysis, 1.63W of heat is removed. It constitute to 64% of maximum heat that can be removed.

### 4.3.3 Comparison of experimental and analytical thermal load for vapor cooled case

Analytical thermal load:
Radiation load                                                                 : 1.96W
Wiring(20 Aluminium wires, each of length 200mm and dia 0.2mm): 0.3W
Glass fibre support structure (LXBXT = 35mmX25mmX1mm)       : 0.1W
Conduction from LN2 tank neck tube                             : 0W
Net thermal load                                               : 2.35W
Experimental heat load:
LN2 hold time(t) from Figure 12                                : 23.3 hrs = 83880 s
Net thermal load                                               = m x $L_{N2}$/ t
                                                               = 2.27 W

m- Mass of liquid nitrogen in the completely filled LN2 tank.
Difference in analytical and experimental heat load is 3%.

### 4.4 Thermal testing of detector assembly

**Detector assembly setup:**

Entire detector assembly(Figure 11c) is enclosed inside aluminium detector housing cooled by copper cold finger from LN2 tank cold plate shown in Figure 11a. Detector assembly and detector housing are thermally isolated from surrounding structures using derlin material for all mechanical support structures. Detector assembly is cooled by small OFHC copper strip from detector housing as shown in Figure 11b. Silicon diodes are used to measure temperature. Below vacuum level of 5x $10^{-5}$ mbar, fill LN2. The temperature values are measured and recorded using Lakeshore Temperature controller and pressure values are also recorded. As the temperature on the entire assembly is stabilised, fill the LN2 tank completely and duration for which temperature is stable gives hold time.

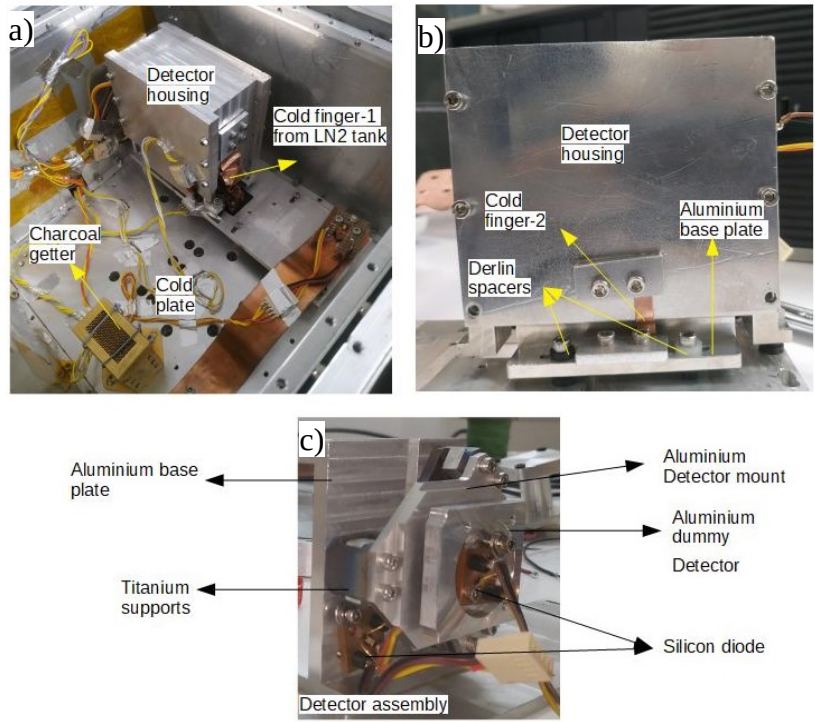

Figure 11: a) Detector housing with cold finger connection from LN2 tank. b) Cold finger from detector housing to detector assembly aluminium base plate. c) Detector assembly with silicon diodes

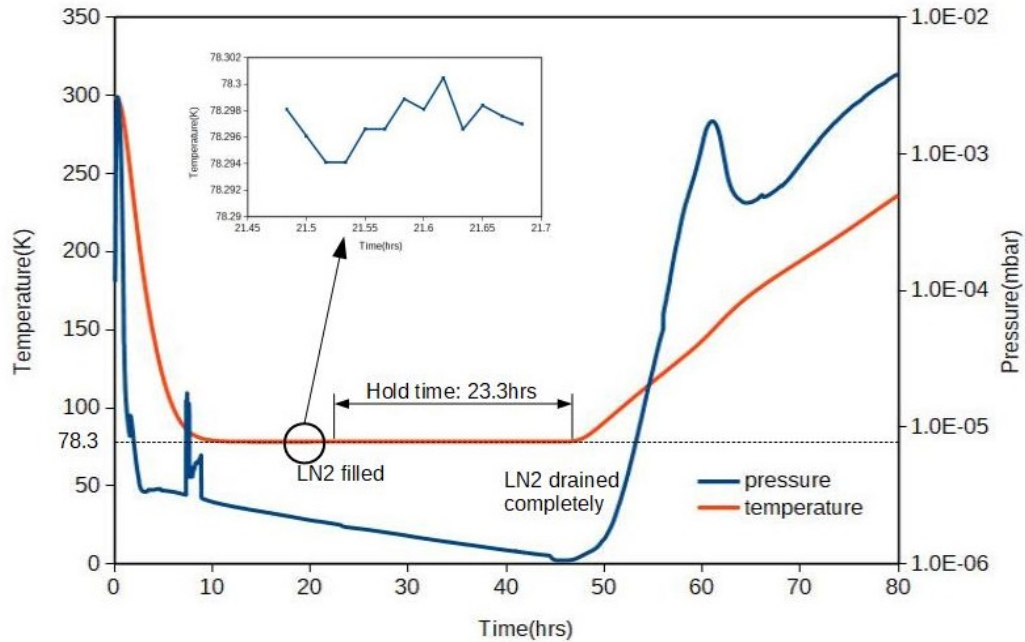

Figure 12: Temperature on dummy detector and Pressure inside the Dewar during cold test

**Results**

Maximum cooling rate : 0.864K/min
Minimum stable temperature (No load): 78.3K at Ahmedabad (76.5K at Gurushikar, Mt.Abu, Rajasthan, India)
LN2 Hold time(hrs) : 23.3hrs
Temperature stability : 0.006K

## SUMMARY

- Analytical thermal heat load estimation of full scale NISP Dewar was done
- Full scale filterwheel assembly and detector-SIDECAR assembly was designed, manufactured and room temperature tests were done
- Test Dewar was designed, manufactured and tested for leaks. Leak rate measured is almost similar to the gas permeation by O-rings. This guarantees leak tightness of the system.
- Charcoal getter working is tested experimentally.
- Experimentally validated analytical heat load calculations of Test Dewar.
- Efficiency of heat transfer from vapor cooling is estimated experimentally. This estimation is used for full scale Dewar analysis.
- Detector assembly is designed, manufactured and tested for required temperature, cooling rate.

## ACKNOWLEDGEMENT

We thank PRL Workshop staff for their efforts in manufacturing the Test Dewar, Filter wheel assembly and Detector assembly.